\documentclass[preprint,floatfix] {revtex4} 
 
\usepackage{graphicx}
\usepackage{subfigure}
\begin{document}

\title{Accurate calculation of the bound states of Hellmann potential}
\author{Amlan K. Roy}
\altaffiliation{Corresponding author. Email: akroy@chem.ucla.edu}
\affiliation{Department of Chemistry and Biochemistry, University of California, 
Los Angeles, CA, 90095-1569, USA}

\author{Abraham F. Jalbout}
\affiliation{Institute of Chemistry, National Autonomous University of Mexico,
Mexico City, Mexico}

\author{Emil I. Proynov}
\affiliation{Q-Chem Inc., 5001 Baum Blvd., Pittsburgh, PA 15213, USA}

\begin{abstract}
Bound states of the Hellmann potential, which is a superposition of the attractive
Coulomb ($-A/r$) and the Yukawa ($Be^{-Cr}/r$) potential, are calculated by using a 
generalized pseudospectral method. Energy eigenvalues accurate up to thirteen to fourteen 
significant figures, and densities are obtained through a nonuniform, optimal spatial 
discretization of the radial Schr\"odinger equation. Both ground and excited states 
are reported for arbitrary values of the potential parameters covering a wide range of
interaction. Calculations have been made for higher states as well as for stronger 
couplings. Some new states are reported here for the first time, which could be useful
for future works. The present results are significantly improved in accuracy over all
other existing literature values and offers a simple, accurate and efficient scheme for
these and other singular potentials in quantum mechanics. 
\end{abstract}
\maketitle

\section{Introduction}
A two-particle system interacting through a combination of the attractive Coulomb 
and the Yukawa potential,
\begin{equation}
v(r)=-A/r + Be^{-Cr}/r
\end{equation}
has received considerable interest for several decades. In this equation, the parameters 
$A$, $B$ characterize the strength of the Coulomb and Yukawa potentials respectively; $C$
is the screening parameter and $r$ signifies the distance between the two particles. 
$A, C$ are positive and $B$ can be both positive as well as negative. Historically, such
a superposed potential with positive $B$ was first studied by Hellmann [1-3] long times
ago and thereafter has been customarily used to include both positive and negative $B$. 
This has found various important applications in the field of atomic and condensed matter
physics; e.g., the electron-core [4,5], electron-ion [6,7], inner-shell ionization [8]
problems, alkali hydride molecules [9], solid-state physics [10,11], etc.

From the theoretical perspectives, this potential has attracted considerable attention
from various workers. Like most other practical physical systems, the corresponding
Schr\"odinger equation (SE) does not offer exact analytical solutions in 
this case too, and one has to resort to the approximate methodologies, such as the 
variational or the perturbative approaches. Some important aspects of the bound-state
spectra of this system are the presence of complex level crossings [12] and the absence 
of accidental degeneracies (characteristics of the pure Coulomb potential). Quite detailed 
calculations were performed [12] using the variational technique including ten 
parameters for wide ranges of the parameters in the potential corresponding to ground, 
as well as low and moderately high values of the $n$ and $\ell$ quantum numbers. Shortly
after that, shifted large N expansion results [13] were reported for these systems which
were more or less of similar accuracy as those of [12], although with limited applicability. 
Besides, 
attempts have been made to use first-order Rayleigh-Schr\"odinger perturbation theory to 
provide approximate analytical formulas for the bound eigenstates [14]. Lately, a combined 
Hellmann-Feynmann theorem and the principle of minimal sensitivity has also been used 
to investigate these states [15]. Analytical formulas for the upper and lower bounds 
[16] have been presented recently by using an envelope method in conjunction with the 
comparison theorem. However, despite all these elegant formalisms, there are several 
problems which deserve more careful and thorough examinations. For example, the prescription 
of [13] yields reasonably good results for very weak screenings and gradually worsens as
$B$ and $C$ increase. For certain other choices of 
the parameters, this leads to divergent energy series for some of the eigenstates. 
Analogous difficulties have also been faced in the recent treatment of these potentials
[15] using the shifted 1/N expansion. It is also worthwhile to note that
although the variational results of [12] were quite accurate and so far have been used
as a standard in the literature for this potential, it would be useful and desirable 
to have more accurate results for these states. The lack of such results in the 
literature is little surprising, especially in the light of the fact that many excellent
and high quality results have been available for both the Coulomb and Yukawa potentials
for many years (see for example [17-19]). Thus a general reliable formalism which can 
offer accurate and physically meaningful results for arbitrary values of the interaction
parameters for both low as well as higher states, would have its own merit.   

Hence it would be of some interest to investigate the spectra of these systems with a 
fresh look. The purpose of this Letter is to employ the generalized pseudospectral (GPS)
method to solve the corresponding SE in a simple and accurate 
manner. To this end, accurate eigenvalues and densities are reported for all the $n \leq
5$ states and the effects of varying the interaction parameters are studied by covering
a large range. The GPS method has emerged as a quite successful formalism to study a 
multitude of atomic and molecular processes in the past years including both electronic
structure and dynamics calculations having Coulomb singularities. Recently it has also been 
shown to be equally successful
for the spiked harmonic oscillators, logarithmic and power-law potentials, the Hulthen and
Yukawa potentials as well as other singular potentials [20-23]. Comparisons with the 
literature data have been made wherever possible. The article is organized as follows: 
Section II gives a brief outline of the GPS method used here to solve the SE in presence 
of the Hellmann potential. A discussion of the results is made in Section III, while we end 
with a few concluding remarks in Section IV.    

\section{The GPS formalism for the solution of Hellmann potential}
\label{sec:method}
This section presents an overview of the GPS formalism along with the mapping procedure used
for solving the radial SE of a Hamiltonian containing a Hellmann potential within the nonrelativistic
framework. Only the essential steps are given and 
the relevant details may be found elsewhere ([20-23] and the references therein). 
Unless otherwise mentioned, atomic units are employed throughout this article. 

The radial SE can be written in the following form, 
\begin{equation}
\left[-\frac{1}{2} \ \frac{\mathrm{d^2}}{\mathrm{d}r^2} + \frac{\ell (\ell+1)} {2r^2}
+v(r) \right] \psi_{n,\ell}(r) = E_{n,\ell}\ \psi_{n,\ell}(r)
\end{equation}
where $v(r)$ is given as in Eq. (1). Here $n$ and $\ell$ signify the usual radial and
angular momentum quantum numbers respectively. 

One of the distinctive features of GPS method is that it allows one to work in a
nonuniform and optimal spatial discretization; a coarser mesh at larger $r$ and a 
denser mesh at smaller $r$, while maintaining a similar accuracy at both the regions. 
Thus it suffices to work with a significantly smaller number of grid points efficiently, 
which is in sharp contrast to some of the commonly used finite difference or finite 
element methods for the singular potentials, where one is almost forced to use 
considerably larger mesh, often presumably because of their uniform nature.   

At the first step a function $f(x)$ defined in the interval $x \in [-1,1]$ is 
approximated by the N-th order polynomial $f_N(x)$ as follows,
\begin{equation}
f(x) \cong f_N(x) = \sum_{j=0}^{N} f(x_j)\ g_j(x),
\end{equation}
which guarantees that the approximation is \emph {exact} at the \emph {collocation 
points} 
$x_j$, i.e.,
\begin{equation}
f_N(x_j) = f(x_j).
\end{equation}
In the Legendre pseudospectral method which we use here, $x_0=-1$, $x_N=1$, and the 
$x_j (j=1,\ldots,N-1)$ are obtained from the roots of the first derivative of the 
Legendre polynomial $P_N(x)$ with respect to $x$, i.e., 
\begin{equation}
P'_N(x_j) = 0.
\end{equation}
The $g_j(x)$ in Eq.~(3) are called the cardinal functions given by,
\begin{equation}
g_j(x) = -\frac{1}{N(N+1)P_N(x_j)}\ \  \frac{(1-x^2)\ P'_N(x)}{x-x_j},
\end{equation}
and satisfy the unique property, $g_j(x_{j'}) = \delta_{j'j}$. At this stage one can 
map the semi-infinite domain $r \in [0, \infty]$ onto the finite domain $x \in [-1,1]$
by the transformation $r=r(x)$. Now introduction of the following algebraic nonlinear
mapping,
\begin{equation}
r=r(x)=L\ \ \frac{1+x}{1-x+\alpha},
\end{equation}
where L and $\alpha=2L/r_{max}$ are the mapping parameters, in conjunction with the  
relation,
\begin{equation}
\psi(r(x))=\sqrt{r'(x)} f(x)
\end{equation}
followed by a symmetrization procedure leads to the transformed Hamiltonian as below, 
\begin{equation}
\hat{H}(x)= -\frac{1}{2} \ \frac{1}{r'(x)}\ \frac{d^2}{dx^2} \ \frac{1}{r'(x)}
+ v(r(x))+v_m(x),
\end{equation}
where $v_m(x)$ is given by,
\begin{equation}
v_m(x)=\frac {3(r'')^2-2r'''r'}{8(r')^4}.
\end{equation}
This has the advantage that one deals with a \emph {symmetric} matrix eigenvalue problem
which can be easily solved by standard available routines to yield accurate eigenvalues
and eigenfunctions. Note that $v_m(x)=0$ for the particular transformation.
 
We have carried out a large number of tests in order to make a detailed check on the
accuracy and reliability of the method by varying the mapping parameters so as to 
produce ``stable'' results with respect to their changes. This procedure was applied
for a variety of potential parameters available in the literature. In this way, a 
consistent set of parameters $\alpha=25, N=200$ and $r_{max}=200$ were chosen which 
seemed to be appropriate for all the calculations performed in this work. The results 
are reported only up to the precision that maintained stability and all our results 
are {\em truncated} rather than {\em rounded-off}. Thus, all the results may be considered
as correct up to the place they are reported. 

\section{Results and Discussion}
First in table 1, we give the computed $2s$ eigenvalues for three values of the parameter 
$B$, \emph{viz., } $0.5, -0.5,$ and $-2$ as a function of the screening parameter $C$
to demonstrate the accuracy of the present calculations. For each $B$, four $C$ values 
have been considered covering both the weak and strong regions. It may be noted here 
that for all the calculations in this work, we assume $A=1$; accordingly our 
$B$ maps to half of the corresponding rescaled parameter of [12], and the computed
eigenvalues obtained in a. u., are half of those of [12]. The results are compared with
the (a) variational results [12], and (b) Rayleigh-Schr\"odinger perturbation calculations
[14]. For $C=0.001$, only the former results are available. It is abundantly clear that the 
present GPS values are significantly better than either of these previously reported results
in all the cases. For all these states, between [12] and [14], former results are seen to be
closer to present values than the latter ones. Usually the variationally calculated 
eigenvalues match up to four to five significant figures with those of ours while the 
perturbation results place these states at higher values in all but one instance ($B=-2$,$C=2$) 
and the accuracy gradually decreases with an increase in the screening parameter $C$. 

\begingroup
\squeezetable
\begin{table}
\caption {\label{tab:table1}Comparison of the calculated negative eigenvalues (in a.u.) 
with the literature for the $2s$ states as functions of $B$ and $C$. Literature results
have been appropriately converted to the current scale of units.}
\begin{ruledtabular}
\begin{tabular}{cccccccc}
$B$    & $C$   & $-$E(This work) & $-$E(Literature) & $B$ & $C$ & $-$E(This work) & 
$-$E(Literature) \\ \hline
0.5    & 0.001 & 0.03174701400990 & 0.031745\footnotemark[1]                     &
0.5    & 0.005 & 0.03367675354994 & 0.033675\footnotemark[1]$^,$\footnotemark[2] \\
0.5    & 2     & 0.11290716132278 & 0.112905\footnotemark[1],0.11115\footnotemark[2]  &
0.5    & 10    & 0.12339007950313 & 0.12289\footnotemark[1],0.123285\footnotemark[2]  \\ 
$-0.5$ & 0.001 & 0.2807509984473  & 0.28075\footnotemark[1]                       &
$-0.5$ & 0.005 & 0.2787748073142  & 0.278775\footnotemark[1],0.27877\footnotemark[2]  \\
$-0.5$ & 2     & 0.1406129511670  & 0.14061\footnotemark[1],0.13943\footnotemark[2]   &
$-0.5$ & 10    & 0.1268366598878  & 0.126835\footnotemark[1],0.126715\footnotemark[2] \\  
$-2$   & 0.001 & 1.1230019984462  & 1.12300\footnotemark[1]                        &
$-2$   & 0.005 & 1.1150498066913  & 1.115050\footnotemark[1],1.115035\footnotemark[2] \\
$-2$   & 2     & 0.2010044938456  & 0.201005\footnotemark[1],0.20219\footnotemark[2]  &
$-2$   & 10    & 0.1342619146710  & 0.13424\footnotemark[1],0.13187\footnotemark[2] \\
\end{tabular}
\end{ruledtabular}
\footnotetext[1] {Ref. [12].}
\footnotetext[2] {Ref. [14].}
\end{table}
\endgroup

Next, the calculated binding energies ($-E_{n,\ell}$) of some of the lowest lying $1s-5g$ states
below $n \leq 5$ are presented for the strongly repulsive ($B=+5$) and strongly attractive
($B=-5$) Yukawa potentials in tables 2 and 3 respectively as a function of the screening
parameter $C$. Four values of $C$ have been considered in both cases, \emph{viz.,} 0.01, 
1, 10 and 100 which essentially covers both the weak and strong regions. $1s-4f$ states 
were studied for a large number of $C$ values ranging from 0.001--10 by [12] while from 
0.05--10 by [13] and these are appropriately quoted here. To our knowledge, no reference 
values are available for $C>10$, and we report here some results in the very high screening
regions ($C=100$). It is noticed that, the present GPS eigenvalues are much superior to 
both the earlier reported values for all these states. For the repulsive Yukawa potential 
case, the variational results [12] are seen to match with our results in the moderate 
screening regions ($C=1,10$); whereas, the accuracy in their result deteriorates for smaller
$C$ (0.01).
Also the low-$\ell$ states deviate more than the high-$\ell$ states. Furthermore, within a 
particular $\ell$, the errors increase as the radial quantum number $n$ increases. Thus the 
$1s, 2s, 3s$ and $4s$ states for $C=0.01$ are in error by 0.54, 3.76, 11.25 and 23.25\% 
respectively. However, for the attractive Yukawa potentials, they show similar kind of 
accuracies and agreements with the present results for both $C=0.01$ and 10. Once again, 
the variational results [12] are seen to be somewhat better than the perturbation results
[13]. For the $1s$ and $2s$ states of $B=+5$ and $B=-5$, appreciable deviations are noticed
in the calculations of [13] from ours ($\approx$29.26 and 25.06\% respectively); 
other $s$ states also suffer more compared to the $p$, $d$ or $f$ states. In both these cases, 
however, one regains the hydrogen-like spectrum in the limit of $C \rightarrow \infty$ 
(100 in the tables), as expected.      

\begingroup
\squeezetable
\begin{table}
\caption {\label{tab:table2}Comparison of the calculated negative eigenvalues (in a.u.) 
with the literature for $B=+5$ as functions of the screening parameter $C$. Literature
results have been appropriately converted to the current scale of units.}
\begin{ruledtabular}
\begin{tabular}{cllll}
State & $C=0.01$        & $C=1$           & $C=10$          & $C=100$  \\ \hline
$1s$ & 0.002362763418(0.00235\footnotemark[1]) & 0.1393937847772(0.139395\footnotemark[1]) 
     & 0.4219751601088(0.421975\footnotemark[1],0.29851\footnotemark[2] 
     & 0.4981833709122   \\
$5s$ & 0.001525033897  & 0.0144970380925 & 0.0193116714697 & 0.0199854358123   \\
$5d$ & 0.001862762081  & 0.0195295800293 & 0.0199999884584 & 0.0199999999999   \\
$4f$ & 0.002300761996(0.00229\footnotemark[1]) & 0.0312056245649(0.031205\footnotemark[1]) 
     & 0.0312499999917(0.03125\footnotemark[1],0.03125\footnotemark[2]) 
     & 0.0312500000000   \\
$5f$ & 0.002054101204  & 0.0199657840037 & 0.0199999999929 & 0.0200000000000   \\
$5g$ & 0.002260639328  & 0.0199992848683 & 0.0199999999999 & 0.0199999999999   \\ 
\end{tabular}
\end{ruledtabular}
\footnotetext[1] {Ref. [12].}
\footnotetext[2] {Ref. [13].}
\end{table}
\endgroup

\begingroup
\squeezetable
\begin{table}
\caption {\label{tab:table3}Comparison of the calculated negative eigenvalues (in a.u.) 
with the literature for $B=-5$ as functions of the screening parameter $C$. Literature
results have been appropriately converted to the current scale of units.}
\begin{ruledtabular}
\begin{tabular}{cllll}
State & $C=0.01$        & $C=1$           & $C=10$         & $C=100$  \\ \hline
$1s$  & 17.95006243069(17.95005\footnotemark[1])  & 13.56679686030    
      & 0.9788396316974(0.97885\footnotemark[1])  & 0.5020427358386   \\
$5s$  &  0.6715273295205                          &  0.0362575479838  
      & 0.0223084737740                           & 0.0200162962261   \\
$5d$  & 0.6714073958450                           & 0.0214355247975
      & 0.0200000117313                           & 0.0200000000000   \\
$4f$  & 1.075741875123(1.07575\footnotemark[1])   & 0.0313020585157
      & 0.0312500000082(0.03125\footnotemark[1],0.0312\footnotemark[2])  
      & 0.0312499999999   \\
$5f$  & 0.6712874380109                           & 0.0200407809113
      & 0.0200000000070                           & 0.0200000000000   \\
$5g$  & 0.6711274566209                           & 0.0200007358716   
      & 0.0200000000000                           & 0.0200000000000   \\
\end{tabular}
\end{ruledtabular}
\footnotetext[1] {Ref. [12].}
\footnotetext[2] {Ref. [13].}
\end{table}
\endgroup

Now we examine the effects of the variation of parameter $B$ on the calculated eigenvalues
up to $1s$ through $5g$. Tables 4 and 5 present such eigenvalues for moderate screenings
($C=0.25$) at four $B$ values ($-10, -1, 1, 100$) and strong screening ($C=1$) at four $B$ 
values ($-25, -10, 1, 10$) respectively. The reference values for these states are quite 
scarce. $n \leq 4$ states of $C=0.25$ were calculated for $B=-10,-1,1$ by [12] while for 
$C=1$, all the $n \leq 5$ states were reported for $B=-25,-10$ only, in the same work [12].
As in the previous tables, the present method offers noticeably improved results for all of
these states compared to both of these. It appears that the variational calculations [12] 
are relatively more accurate in the low screening regions than in the stronger regions. 
   
\begingroup
\squeezetable
\begin{table}
\caption {\label{tab:table4}Comparison of the calculated negative eigenvalues (in a.u.) 
with the literature for $C=0.25$ as functions of $B$. Literature results have been
appropriately converted to the current scale of units.}
\begin{ruledtabular}
\begin{tabular}{cllll}
State & $B=-10$        & $B=-1$           & $B=1$         & $B=100$      \\ \hline
$1s$ & 58.04198638290(58.042\footnotemark[1])     & 1.771691001196(1.77169\footnotemark[1])
     & 0.1105241235947(0.110525\footnotemark[1])  & 0.0251808092657      \\
$5s$ & 0.7519807159635                            & 0.0271230027804
     & 0.0142921176781                            & 0.0067962156719      \\
$3d$ & 4.498053096472(4.498055\footnotemark[1])   & 0.0862527058258(0.086255\footnotemark[1])
     & 0.0451094116513(0.04511\footnotemark[1])   & 0.0221116573287      \\
$5d$ & 0.7042645932150                            & 0.0240731854698
     & 0.0177686031869                            & 0.0105703836625      \\
$5f$ & 0.6551355037045                            & 0.0217548871225
     & 0.0189834793491                            & 0.0131097965268      \\
$5g$ & 0.5870275279097                            & 0.0203601943459      
     & 0.0197229049653                            & 0.0161504601523      \\
\end{tabular}
\end{ruledtabular}
\footnotetext[1] {Ref. [12].}
\end{table}
\endgroup

\begingroup
\squeezetable
\begin{table}
\caption {\label{tab:table5}Comparison of the calculated negative eigenvalues (in a.u.) 
with the literature for $C=1$ as functions of $B$. Literature results have been 
appropriately converted to the current scale of units.}
\begin{ruledtabular}
\begin{tabular}{cllll}
State & $B=-25$        & $B=-10$           & $B=1$         & $B=10$      \\ \hline
$1s$ & 313.7035561801(313.7035\footnotemark[1])  & 51.14471457780(51.14245\footnotemark[1])
     & 0.2562317633033                           & 0.1170817257811       \\
$5s$ & 0.9549949909076(0.9550\footnotemark[1])   & 0.0619515214000(0.0615\footnotemark[1])
     & 0.0175035546929                           & 0.0135892847495       \\
$5d$ & 0.5235991105174(0.5236\footnotemark[1])   & 0.0324039077933(0.0323\footnotemark[1])
     & 0.0198787392313                           & 0.0192456899035       \\
$5f$ & 0.1025542046267(0.10175\footnotemark[1])  & 0.0200939350152(0.0201\footnotemark[1])
     & 0.0199927279413                           & 0.0199358878025       \\
$5g$ & 0.0200040154394(0.0200\footnotemark[1])   & 0.0200014959681(0.0200\footnotemark[1])
     & 0.0199998554056                           & 0.0199985879641       \\
\end{tabular}
\end{ruledtabular}
\footnotetext[1] {Ref. [12].}
\end{table}
\endgroup

Now fig. 1 shows the variation of the radial probability distribution functions with respect
to the interaction parameters $B$ and $C$; (a) shows this for $C=1$ at four $B$ values, 
\emph{viz.,} $-0.1,1,10$ and 100 while (b) shows this for $B=0.5$ at four $C$ values, 
\emph{viz.,} 0.01, 0.1, 1 and 10 respectively. It is seen that with an increase of $B$, 
the density distribution oozes out to the larger values of $r$ and the peak values get 
reduced, while exactly opposite behaviour is observed as the parameter $C$ is increased. 
Additionally, the radial density distributions of the Hellmann potential ($B=1, C=10$) 
in fig. 2 shows the desired number nodes and peaks for the first three states in (b), (c)
and (d) corresponding to $\ell=0,1$ respectively, along with the potential in (a). 

\begin{figure}
\centering
\begin{minipage}[t]{0.4\textwidth}\centering
\includegraphics[scale=0.35]{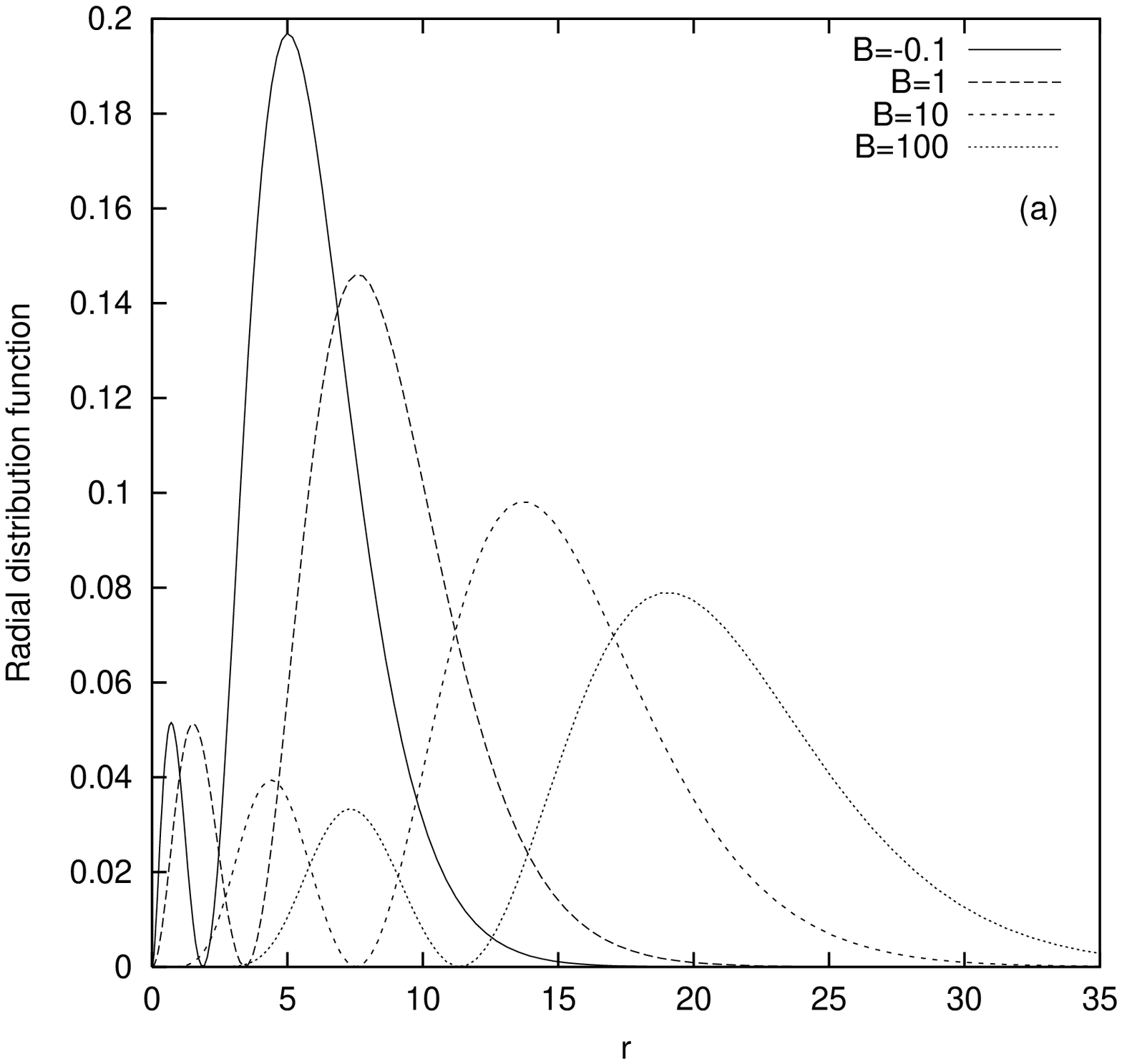}
\end{minipage}
\hspace{0.15in}
\begin{minipage}[t]{0.35\textwidth}\centering
\includegraphics[scale=0.35]{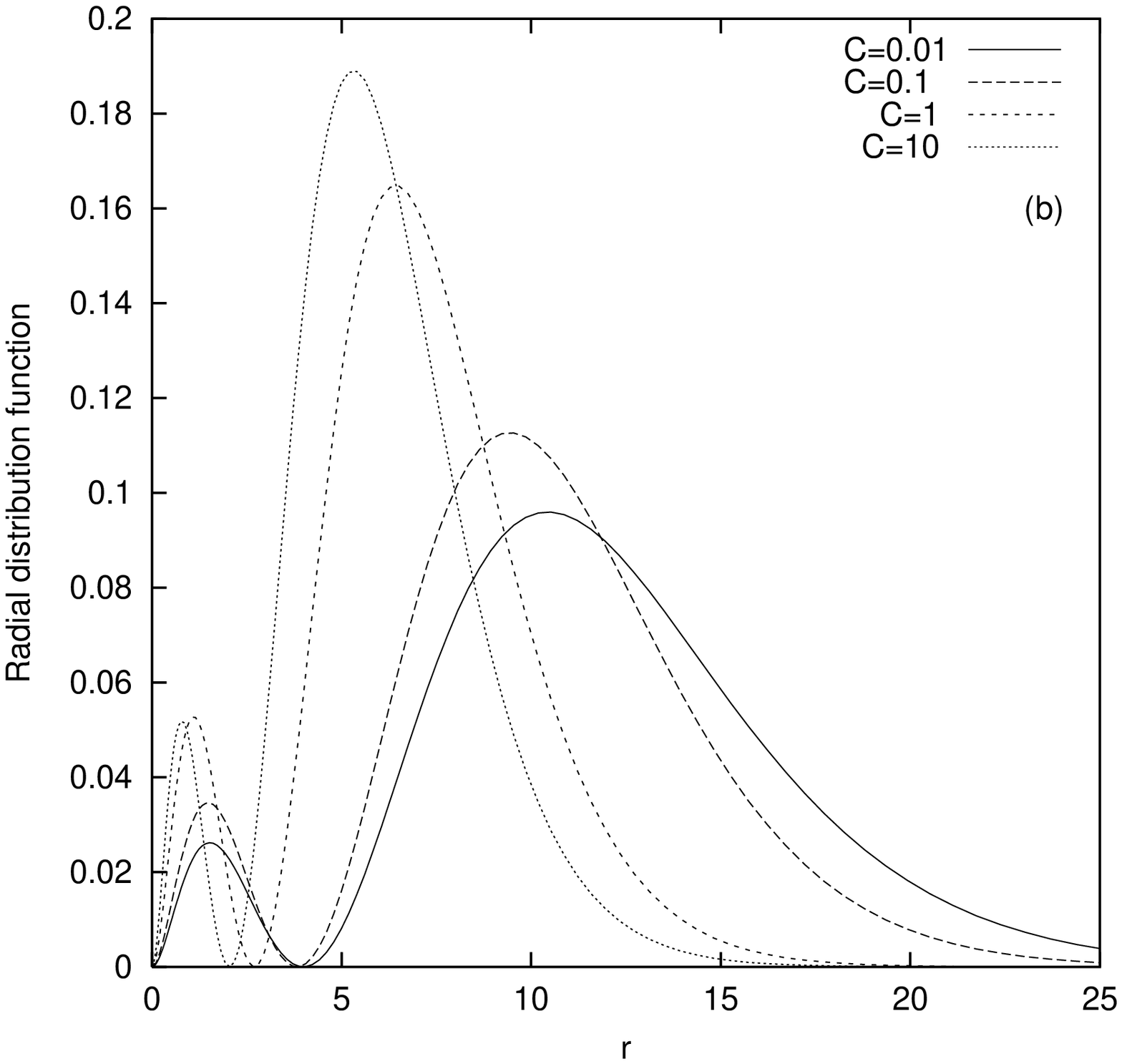}
\end{minipage}

\caption[optional]{Variation of the radial probability distribution function, 
$|rR_{n,\ell}|^2$, for $2s$ state of the Hellmann potential with respect to the
parameters $B$ and $C$. (a) $B=-0.1,1,10,100$ and $C=1$; (b) $B=0.5$ and $C=0.01, 0.1,
1,10$ respectively.}
\end{figure}

\begin{figure}
\centering
\begin{minipage}[t]{0.4\textwidth}\centering
\includegraphics[scale=0.35]{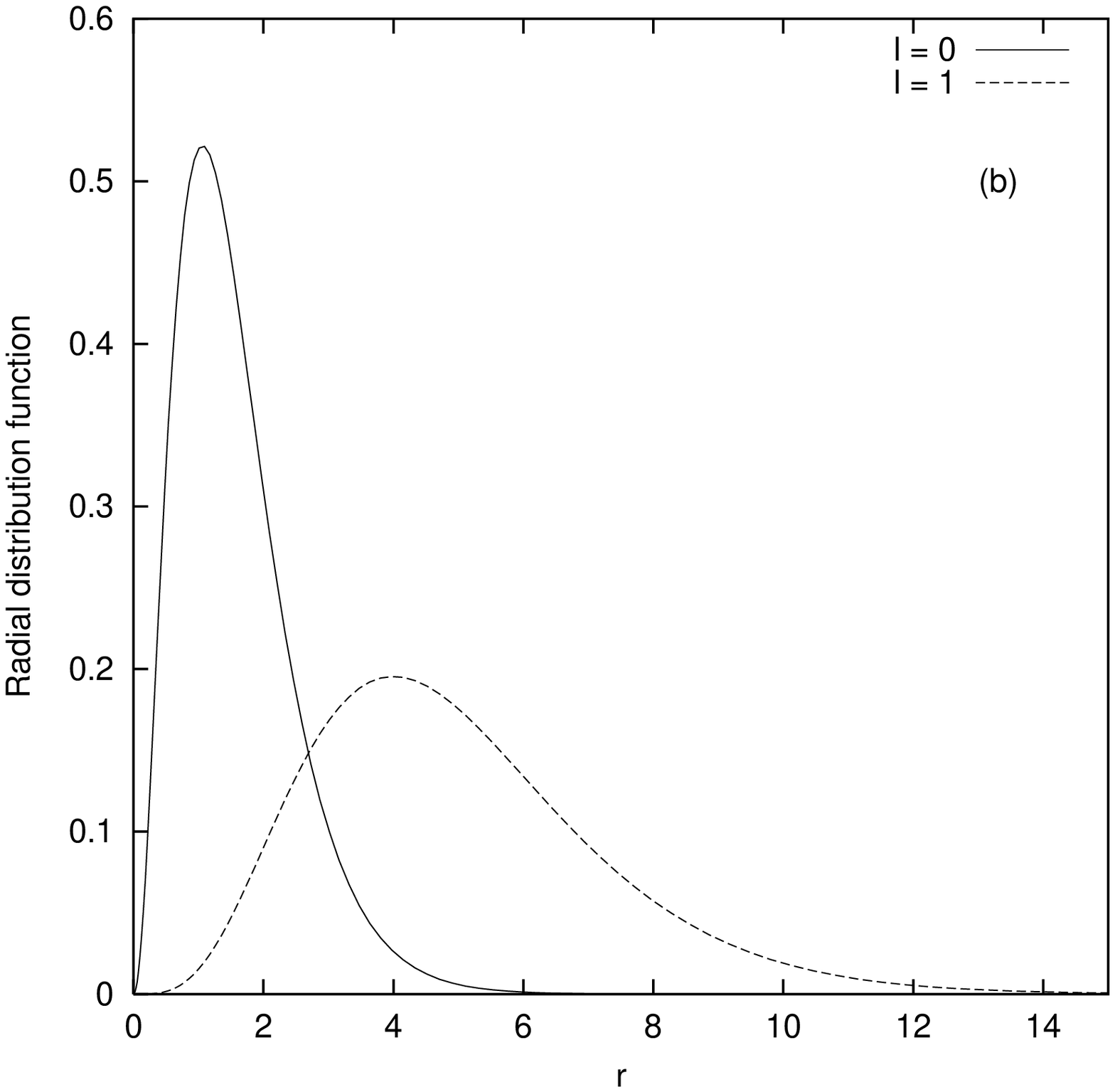}
\end{minipage}
\hspace{0.15in}
\begin{minipage}[t]{0.35\textwidth}\centering
\includegraphics[scale=0.35]{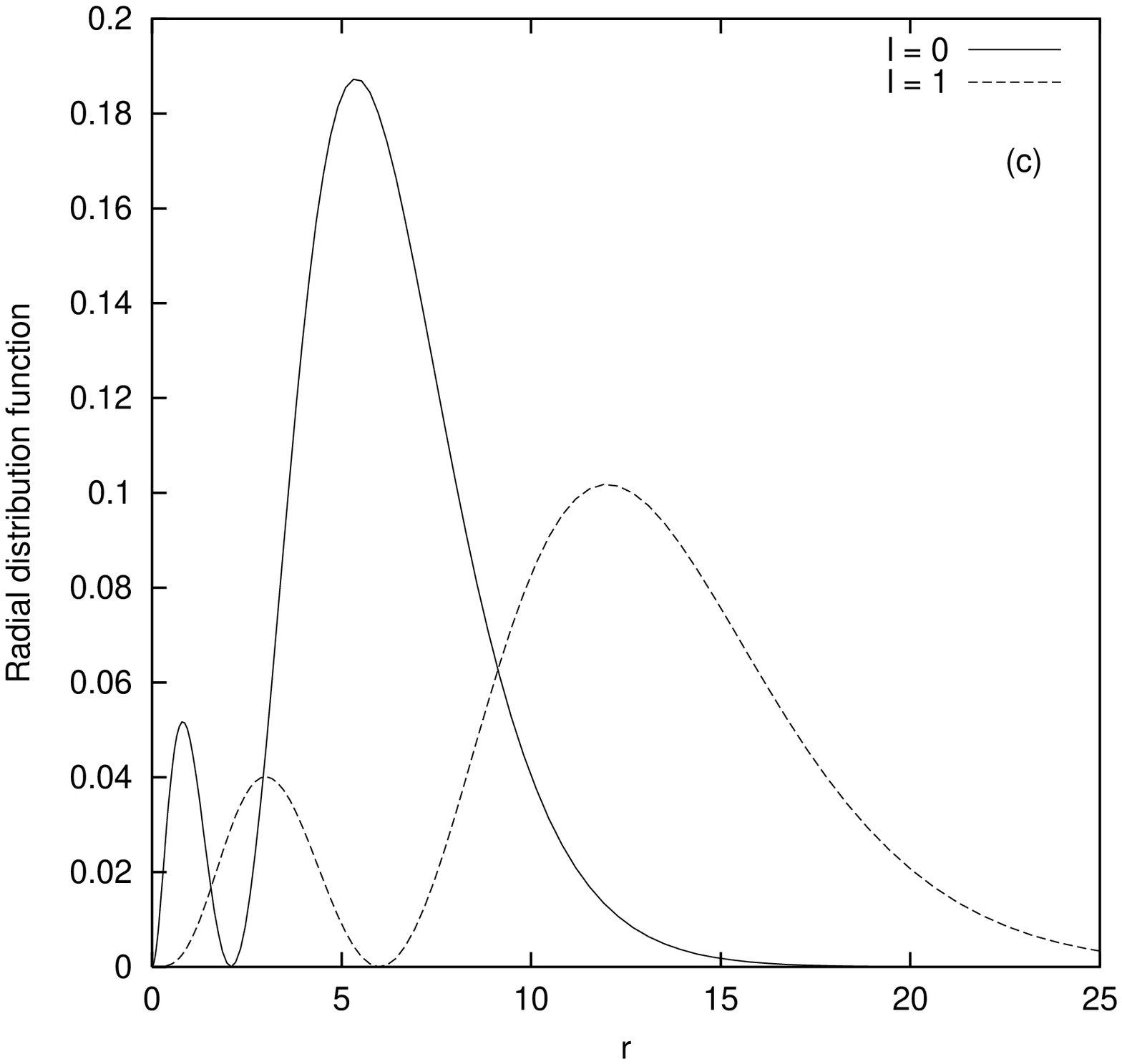}
\end{minipage}
\\[30pt]
\begin{minipage}[b]{0.4\textwidth}\centering
\includegraphics[scale=0.35]{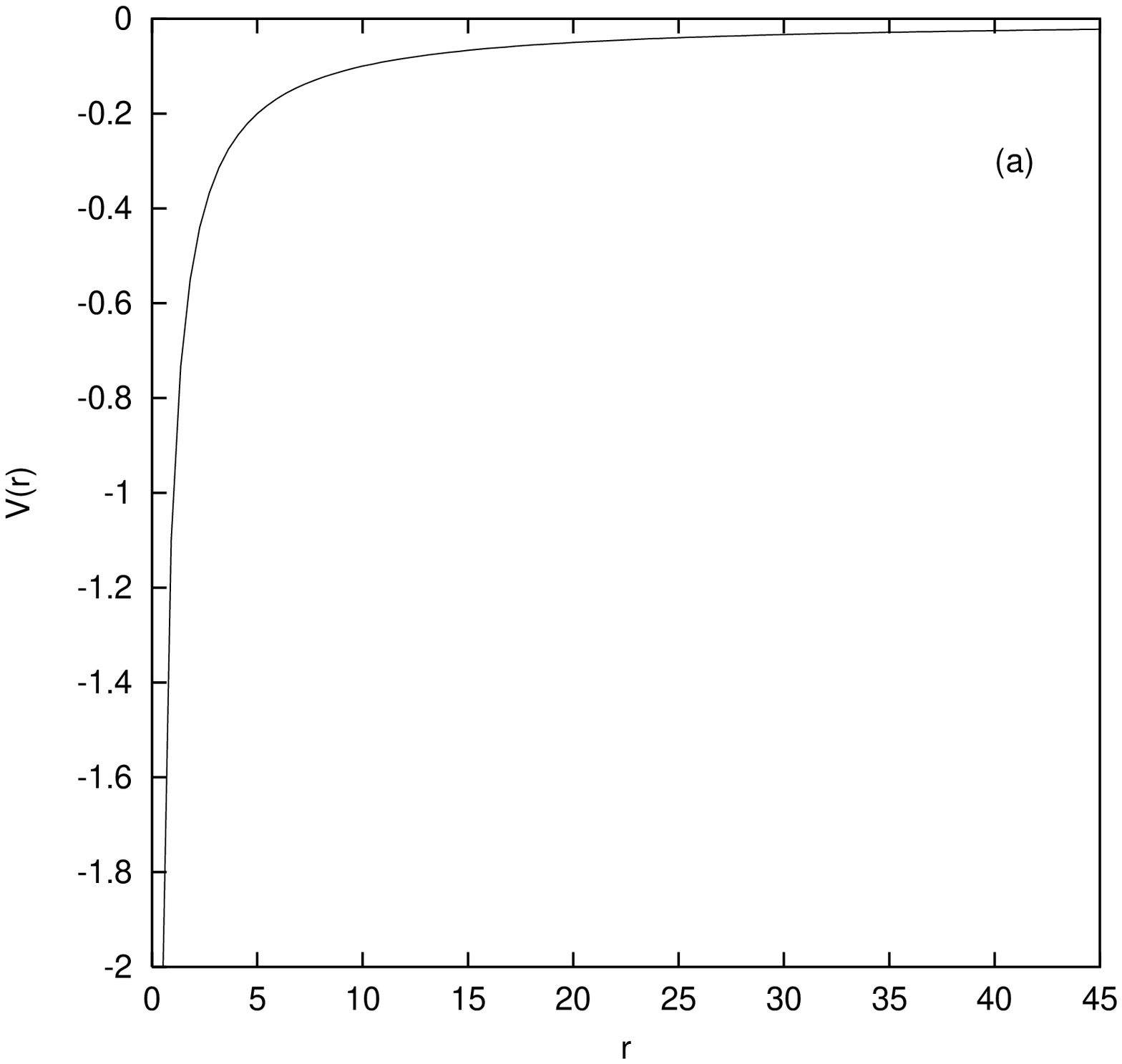}
\end{minipage}
\hspace{0.15in}
\begin{minipage}[b]{0.35\textwidth}\centering
\includegraphics[scale=0.35]{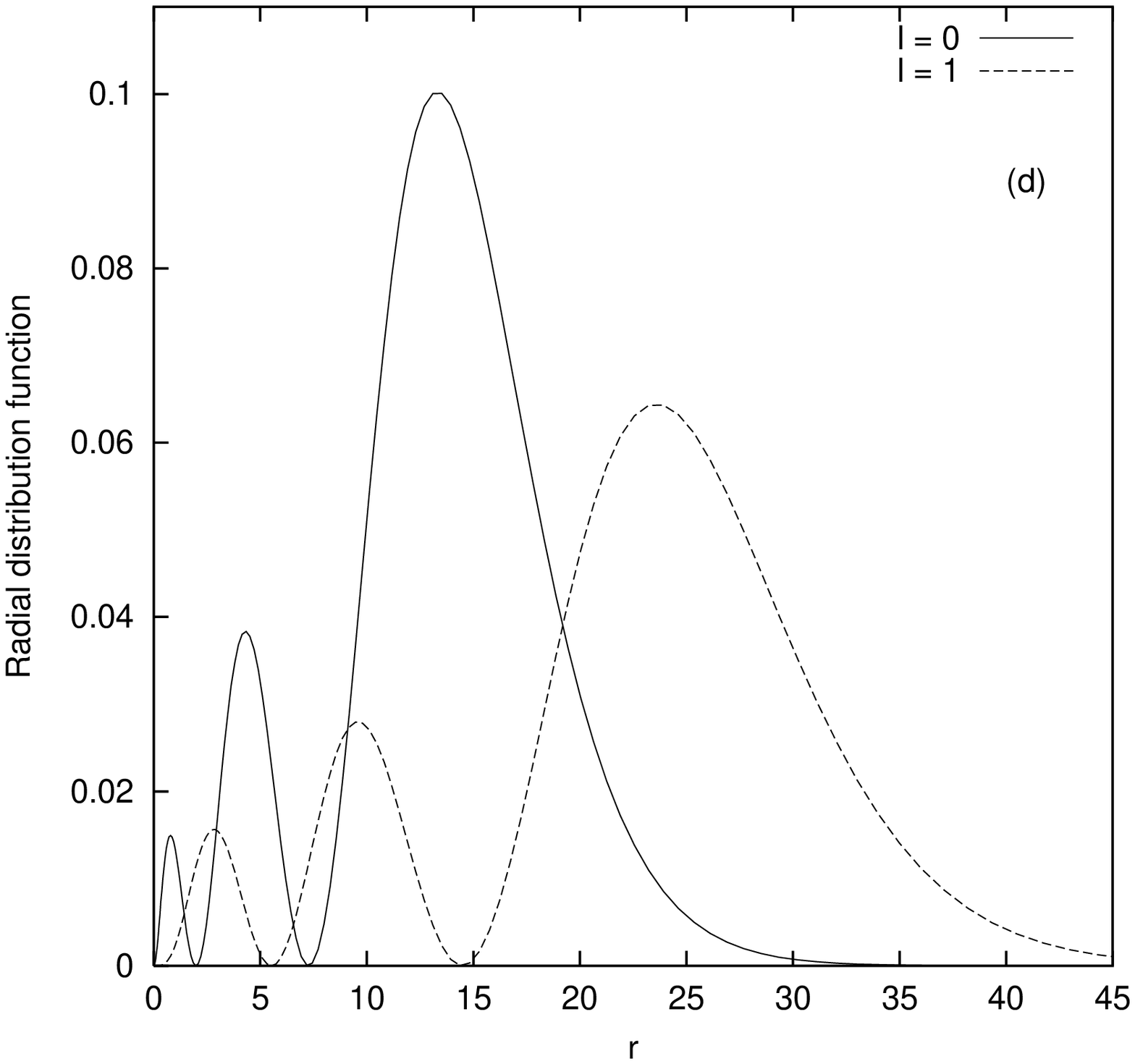}
\end{minipage}
\caption[optional]{Variation of the radial probability distribution functions, 
$|rR_{n,\ell}|^2$, for the first three states corresponding to $\ell=0,1$ of the Hellmann
potential with $B=1$ and $C=10$. (a) the potential, (b) ground state, (c) first excited
state  and (d) second excited state.}
\end{figure}

Before passing, a few comments may be made. It is worth mentioning that although many 
attractive and elegant formalisms have been proposed over the years, it is usually 
quite a difficult task to achieve faster convergence and high accuracy results for the 
singular potentials at the same time by using the standard finite difference methods. As
one author pointed out [24], a six- or seven-decimal place accuracy for the harmonic 
potential including an inverse quartic and sextic anharmonicity required at least 80,000
radial grid points for some of the lower states. The GPS method, employed in the current 
work possesses the simplicity of the finite difference and/or the finite element methods
while simultaneously retaining the attractive features of the basis-set variational 
formalisms such as high accuracy and fast convergence. As already mentioned, all the  
calculations in this work have been done with only 200 radial grid points. 
   
\section{Conclusion}
Discrete bound-state spectra of the superposed Coulomb and Yukawa potentials, are studied 
in detail by accurately calculating the eigenvalues and densities through the GPS method.
The formalism is simple, computationally efficient, reliable and accurate. Low as well as 
high states are calculated for arbitrary values of the interaction parameters covering 
weak and strong couplings with equal ease and accuracy. For all the 15 states belonging to
$n \leq 5$, the present method offers results which significantly improves all other 
hitherto reported literature values. In view of the simplicity and accuracy offered by this
method for the physical systems studied in this work, it might have equally successful 
and fruitful applications in various other branches of quantum mechanics. 

\begin{acknowledgments}
AKR thanks Professors D. Neuhauser and S. I. Chu for useful discussions. He acknowledges the warm 
hospitality provided by the Univ. of California, Los angeles, CA, USA. EP gratefully 
acknowledges Q-Chem Inc., for support.  
\end{acknowledgments}

\end{document}